# BIASED BY DESIGN: LEVERAGING AI BIASES TO ENHANCE CRITICAL THINKING OF NEWS READERS

*Completed Research Paper*

Liudmila Zavolokina, University of Lausanne and University of Zurich, Lausanne and Zurich, Switzerland, liudmila.zavolokina@unil.ch

Kilian Sprenkamp, University of Zurich, Zurich, Switzerland, kilian.sprenkamp@uzh.ch

Zoya Katashinskaya, University of Zurich, Zurich, Switzerland, zoya.katashinskaya@uzh.ch

Daniel Gordon Jones, University of Zurich, Zurich, Switzerland, danielgordon.jones@uzh.ch

## Abstract

*This paper explores the design of a propaganda detection tool using Large Language Models (LLMs). Acknowledging the inherent biases in AI models, especially in political contexts, we investigate how these biases might be leveraged to enhance critical thinking in news consumption. Countering the typical view of AI biases as detrimental, our research proposes strategies of user choice and personalization in response to a user's political stance, applying psychological concepts of confirmation bias and cognitive dissonance. We present findings from a qualitative user study, offering insights and design recommendations (bias awareness, personalization and choice, and gradual introduction of diverse perspectives) for AI tools in propaganda detection.*

## 1 Introduction

Throughout history, propaganda has posed a significant challenge, impacting societies and the functioning of democracies. With the rise of digital channels, the volume and speed at which propaganda is disseminated increased. With this speed, many individuals consume information indiscriminately, making them unable to discern propaganda and biases in the news, thus becoming more susceptible to them. Moreover, recently, governments and political actors across the globe, operating in both democratic and autocratic regimes, have started increasingly using generative artificial intelligence (AI) to create texts, images, and videos (Ryan-Mosleyarchive, 2023). These AI-generated materials are often created to manipulate public opinion in favour of specific agendas (Ryan-Mosleyarchive, 2023). Computational methods can detect propaganda effectively (Ahmed et al., 2021; Della Vedova et al., 2018; Qawasmeh et al., 2019), offering a countermeasure against the spread of misleading and manipulative information. For example, prior studies examined how propaganda could be detected in a document (Barrón-Cedeno et al., 2019) or at the sentence level (Da San Martino, Cresci, et al., 2020; Da San Martino, Shaar, et al., 2020; Piskorski et al., 2023). In recent research, Large Language Models (LLMs), a class of pre-trained AI models with human-level language capabilities of comprehension and generation of text, are increasingly used for propaganda and fake news detection (Hasanain et al., 2024; Patil et al., 2024; Sprenkamp et al., 2023) due to their benefits of advanced natural language understanding and efficient processing of large datasets.

However, one well-known problem is that AI models are not infallible and carry on their own inherent biases stemming from the data they were trained on (Mehrabi et al., 2021). In different contexts, such biases may result in undesirable outcomes such as reinforcing discrimination or increasing economic disparities (Min, 2023). In propaganda detection, these biases, especially political ones, can significantly influence the result of detecting and explaining propaganda to users. The task of propaganda detection is subjective and complex making the pursuit of an objective' ground truth' particularly challenging.

LLMs are particularly biased due to their training on vast, unfiltered internet corpora, which often contain biased or unbalanced viewpoints. Despite significant efforts, completely eliminating biases from AI models is not only challenging but can also inadvertently lead to the creation of new types of biases (Gallegos et al., 2023). Since LLMs are pretrained and do not necessarily require additional training, they can be used conveniently as-is. However, this convenience presents a trade-off: one must contend with the biases inherent in LLMs. Consequently, this requires new approaches on how these inherent biases can be dealt with across different application fields, thus, motivating our paper.

While eliminating these biases entirely is challenging, we propose a novel perspective: instead of viewing AI biases solely as 'bad', we explore how they can be leveraged to enhance users' critical thinking. In particular, we investigate how understanding the political biases inherent in AI models, especially those relevant to propaganda detection, can encourage readers to critically evaluate information rather than passively accept it. Drawing on psychological theories such as confirmation bias and cognitive dissonance, we explore how propaganda detection systems can be designed to either align with or challenge users' political viewpoints. We formulate the following research question: *How can we leverage the inherent political biases of LLMs in the design of AI propaganda detection tools to enhance users' critical thinking?*

We propose two strategies to achieve this: (i) providing users with a choice of AI models based on their political preferences, and (ii) personalizing the system's responses to either reinforce or challenge the user's political stance. These strategies aim to stimulate deeper critical engagement with news content, ultimately empowering users to become more critical consumers of information. Our findings from qualitative interviews provide insights into user perceptions of these approaches and suggest design recommendations for AI tools in propaganda detection. By acknowledging and purposefully utilizing inherent AI biases, our work contributes to ongoing discussions about how AI systems can be designed to both detect propaganda and promote critical thinking.

## 2    Case description

This work builds upon prior research on designing a propaganda detection tool named Apollolytics that nudges its users, i.e., readers of digital news, to think more critically about the news content they consume while reading it (Hoferer et al., 2025; Sprenkamp et al., 2023; Zavolokina et al., 2024). This section provides more context about the propaganda detection tool.

The propaganda detection tool is designed as a browser extension, integrating into the user's online news-reading experience (see Figure 1, left). As users navigate through various news sources and read news articles, they either select a piece of text to be scanned or Apollolytics automatically scans the content of the webpage (if turned on), highlighting statements identified as propaganda techniques with distinct colors. These colors serve as visual cues alerting the user on-the-fly about statements that may include propagandistic or manipulative content. When the user hovers over the highlighted text, they see a concise yet informative explanation of why this statement was identified as a propaganda technique. This feature not only alerts users but educates them in recognizing propaganda techniques in news. Prior research has demonstrated that such a tool does, to a certain extent, enhance news readers' critical thinking and propaganda awareness (Zavolokina et al., 2024).

We demonstrate the high-level architecture and the process of how the Apollolytics extension works to detect propaganda techniques and create explanations in Figure 2. In a nutshell, the extension sends the text of interest accompanied by a specific prompt (Zavolokina et al., 2024) to the OpenAI API. From the text, OpenAI's GPT-4 model identifies which propaganda techniques are found in the given article. Subsequently, the model is prompted to create explanations for the previously identified propaganda techniques and mark where the given technique can be found in the text.

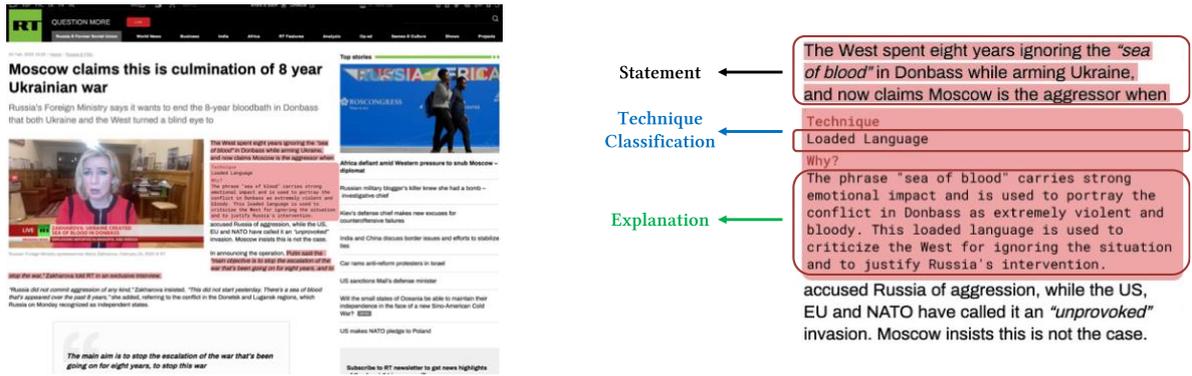

*Figure 1.* Interaction with Apollolytics: the browser extension with propaganda technique detection and explanation (left) and the three important elements of Apollolytics: (i) the statement, classification of (ii) the detected propaganda technique, and (iii) the explanation (right).

These results, i.e. (i) the statements, (ii) identified propaganda techniques, and (iii) explanations, are returned to the browser extension which localizes and visually marks these propaganda techniques and their explanation in the text. In this process, potential political biases of the model can influence all three elements (shown in Figure 1, right): the classification of particular (i) statements, (ii) the propaganda techniques, and the (iii) explanations. For example, a model with a bias towards a particular political viewpoint might disproportionately identify statements aligning with the opposite viewpoint as propaganda, apply certain propaganda techniques more frequently to such statements, and generate explanations that reflect this bias. For simplicity and clarity, this paper primarily concentrates on biases in explanations.

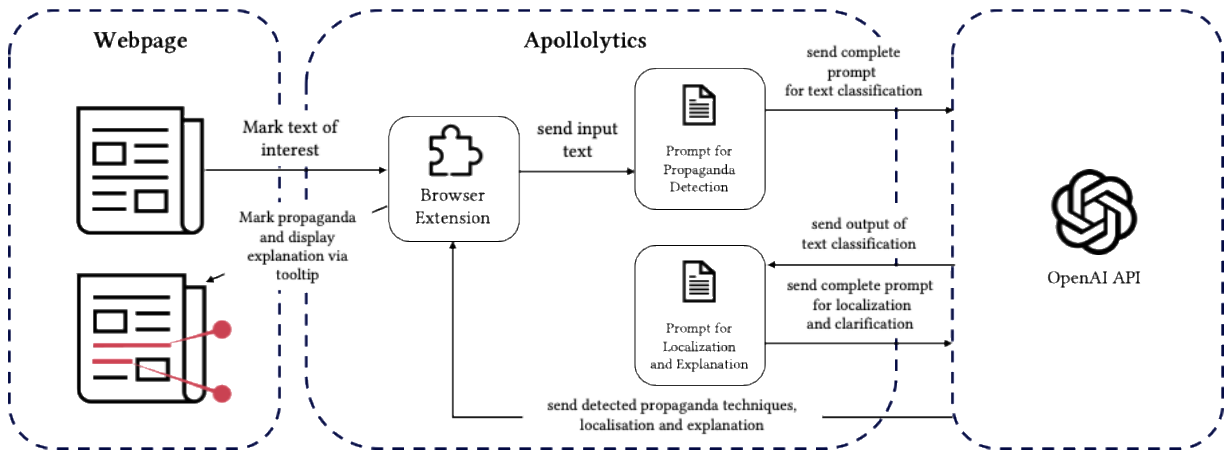

*Figure 2.* High-level architecture and the process flow of Apollolytics.

## 3 Related Work

### 3.1 Biases in AI

Biases in AI models are a widely recognized and well-researched issue (Bender et al., 2021; Mehrabi et al., 2021). Examples of biases in AI models are gender bias (Dastin, 2022), where algorithms might favour one gender over another, and racial bias, leading to different treatment of certain ethnic groups (Kleinberg et al., 2016; Min, 2023) and political bias (Rozado, 2023), which involves a model exhibiting preferential treatment or discrimination toward specific political ideologies or parties. Such political bias can manifest in how information is selected, framed, or ranked by an AI system, thereby influencing the visibility of certain viewpoints or shaping public opinion (Rozado, 2023).. These biases occur due to various factors, including skewed or non-representative training data, inherent biases in algorithm design, and feedback loops in user interactions (Mehrabi et al., 2021). In the case of data, biases may stem from how data is collected and processed or from historical and societal prejudices embedded in the data. Algorithmic biases can arise independently from the nature of the data, influenced by the objectives, features, and methodologies encoded in the algorithm's design. Lastly, biases in user interaction can occur, as how users engage with AI systems can influence future data collection and algorithm development, creating a self-reinforcing cycle of bias.

These self-reinforcing cycles can lead to so-called "echo chambers," where users are continuously presented with information that aligns with their existing views (Quattrociocchi et al., 2016). Over time, this homogenous content ecosystem may limit the diversity of perspectives they encounter, reinforcing existing biases. Moreover, as AI systems adapt to user preferences, they may unintentionally perpetuate polarized online environments (Ohagi, 2024). Such echo chambers can exacerbate political bias by amplifying certain ideologies while minimizing or excluding others (Levy & Razin, 2019). Consequently, user interactions can have a significant impact on the broader information landscape, further entrenching bias in AI-driven platforms.

LLMs are particularly vulnerable to biases, a problem exacerbated by their self-supervised pre-training approach (Bender et al., 2021). In this approach, an LLM is trained on vast amounts of data, which may include biased content. Consequently, the biases inherent in this pre-trained data will likely be reflected in the model's outputs. Researchers often utilize these pre-trained models as a foundation to achieve state-of-the-art results for their respective downstream tasks, thus inheriting the bias of the foundation model (Bender et al., 2021; Bommasani et al., 2021).

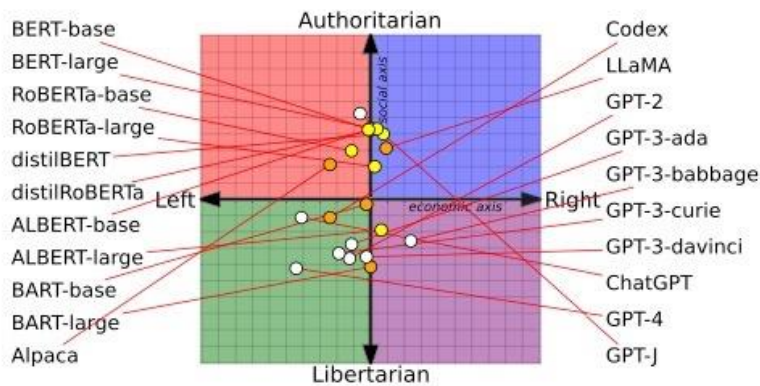

*Figure 3.      Political leaning of various language models (Feng et al., 2023).*

Current studies on the political biases in LLMs such as BERT (Devlin et al., 2019), RoBERTa (Liu et al., 2019), GPT (Brown et al., 2020; OpenAI, 2023; Radford et al., 2019), or Llama (Touvron et al., 2023) reveal distinct bias patterns when analyzed using standardized tools like the political compass[1] (Feng et al., 2023), as visualized in Figure 3. The authors discovered that older Language Models, such as BERT, exhibit a more conservative bias, which can be attributed to the nature of their training data. During their development phase, these models were predominantly trained on older literary texts, reflecting more conservative views. In contrast, newer language models have been trained on contemporary web-based texts, which tend to embody more liberal perspectives, resulting in a shift towards more liberal-leaning biases (Feng et al., 2023). Moreover, the authors demonstrated that the initial political biases are kept by the model when it is fine-tuned on a downstream task (Feng et al., 2023). These inherited biases are especially important in highly politicized tasks like propaganda detection. Propaganda is often understood as deliberately shaping an individual's political opinions through various rhetorical and psychological strategies (Da San Martino, Shaar, et al., 2020). The perception of propaganda typically carries negative connotations, conjuring images of manipulation and deceit. However, the ethical evaluation of propaganda is not inherently black and white; our perception of propaganda often depends on the alignment of values between the propagandist and the recipient (Walton, 1997). Hence, when detecting propaganda through LLMs, what the model perceives as propaganda depends on the political bias of the model.

---

[1] www.politicalcompass.org/test

Mitigation strategies for biases within LLMs focus on four core strategies: preprocessing, in-training, intra-processing, and post-processing (Gallegos et al., 2023). Preprocessing addresses bias by modifying training data or prompts before any model updates. For instance, "counterfactual data augmentation" swaps gendered terms to balance representation in the dataset (Lu et al., 2020), while carefully selected prompts can reduce hateful or harmful completions (Abid et al., 2021). In-training mitigation introduces fairness-oriented objectives into the model's optimization. These may involve adversarial learning, in which a secondary network attempts to predict protected attributes from model representations so that the main model is penalized and discouraged from encoding such demographic cues (Zhang et al., 2018), or loss regularizers that push the model to equalize how frequently it associates certain demographic groups with specific terms (Qian et al., 2019). Intra-processing methods constrain the model's behavior at inference, without retraining. They alter next-token probabilities (sometimes with specialized classifiers that detect sexism or toxicity) to steer the system away from reproducing biased content (Sheng et al., 2020). Post-processing rewrites fully generated text, typically using a rule-based or neural "editor" to detect offensive or biased phrases and convert them into fairer versions (Wang et al., 2022). Because each strategy carries its own trade-offs in complexity and efficacy, researchers often combine several approaches: for example, using data augmentation to expand coverage of minority examples, and then applying adversarial learning during fine-tuning to minimize residual biases.

However, as multiple factors are the causes of bias within AI models, e.g., training data, inherent biases in algorithm design, and feedback loops in user interactions, it is unlikely that completely unbiased models will ever become available. In the context of propaganda detection, leveraging the inherent bias of a model can be advantageous for system designers. By understanding the political inclinations of the end-user, a model can be selected that either aligns with or opposes the user's viewpoints. This personalization approach may potentially enhance the effectiveness of the propaganda detection tool by aligning it more closely with the user's perspective or by intentionally presenting a contrasting viewpoint.

### 3.2  Cognitive biases and critical thinking

This section explores the relationship between critical thinking and two distinct, yet related psychological concepts of human cognition, i.e., confirmation bias (Kahneman, 2011; Tversky & Kahneman, 1974) and cognitive dissonance (Harmon-Jones & Mills, 2019), that shape our understanding and interpretation of information. For this study, we conceptualize confirmation bias and cognitive dissonance as two extremes on a continuum of user perception when encountering information that either aligns with or contradicts their viewpoints.

The vast amounts of information consumed daily compel us to adopt mental shortcuts or heuristics for efficient information processing. This often leads to scenarios where personal preferences, beliefs, and past experiences heavily influence our agreement or disagreement with the information we encounter. For instance, in the context of news reading, this can manifest as a tendency to favour news sources or articles that align with our existing beliefs while dismissing those that contradict them. This phenomenon is recognized as confirmation bias, a concept originally introduced in the seminal work of Kahneman and Tversky (Kahneman, 2011; Tversky & Kahneman, 1974). Confirmation bias leads to effects like echo chambers and polarization, with people becoming less open to differing opinions (Dahlgren, 2020; Dubois et al., 2020). Online spaces particularly facilitate the creation of these echo chambers, making them more efficient, as individuals can easily find and connect with others who share similar opinions. In HCI research, studies have predominantly focused on identifying human biases, quantifying them and developing strategies to make users aware of and educated about their own biases to mitigate their impact (Boonprakong, Chen, et al., 2023; Boonprakong, Tag, et al., 2023; Ji, 2023; Thornhill et al., 2019). However, completely eliminating human biases is hardly possible as they are a natural part of human cognition (Kahneman, 2011).

While confirmation bias leads us to favour information that aligns with our beliefs, cognitive dissonance occurs when our beliefs are confronted with conflicting information (Harmon-Jones & Mills, 2019). If this dissonance is not alleviated by changing these beliefs, it can lead to misperceiving or misinterpreting the information, rejecting or refuting it, seeking agreement from like-minded individuals, and trying to persuade others to accept these beliefs (Harmon-Jones & Mills, 2019). Cognitive dissonance has been studied across domains such as marketing (Telci et al., 2011), psychology (Bran & Vaidis, 2020), economics (Akerlof & Dickens, 1982), and behavioral finance (Olsen, 2008). In HCI, cognitive dissonance is primarily seen as a negative factor to be minimized (Im et al., 2014; Nayak et al., 2021) or as a particular mental state of users as a result of interaction, for example, in contexts like interactions with robots or other systems (Levin et al., 2013; Xin et al., 2017).

When confronted with information, like an explanation of a propaganda technique, a user's cognitive response lies on a spectrum between confirmation bias and cognitive dissonance. This response is influenced by their degree of agreement and their previous beliefs and experiences. Prior research, e.g. (Boonprakong, Chen, et al., 2023; Richter & Maier, 2017; Schwind et al., 2011), has demonstrated that for certain contexts, presenting information contrary to the user's views may indeed be beneficial for achieving the intended outcome, such as reducing confirmation bias. For example, in information search, preference-inconsistent recommendations reduce confirmation bias (Schwind et al., 2011).

## 4   Leveraging Biases in Propaganda Detection

In this section, we propose scenarios aimed at enhancing users' critical thinking, along with strategies and their technical implementations that could facilitate these outcomes. Contrary to the conventional belief that cognitive biases hinder critical thinking, our research explores the idea that combining model biases (such as models' political leaning, relevant for the context of news) with human cognitive biases can actually improve it. We suggest measuring the difference between a user's political viewpoint and the political leaning of LLMs (for example, measured by the political compass test), referred to as 'opinion difference' (see Figure 4). If a user holds an opinion and the explanation generated by the LLM aligns with it, this represents a low opinion difference. Similarly, if the user has an opinion and the explanation provided contrasts with it, this indicates a high opinion difference.

Our research poses a question of whether the explanations should align with a user's political beliefs and viewpoints ('confirmation bias' scenario), oppose them ('cognitive dissonance' scenario), or should reflect the middle ground (as a representation of neutrality scenario) to promote critical analysis of an article (see Figure 5 with the possible scenarios visualized and Table 1 for description).

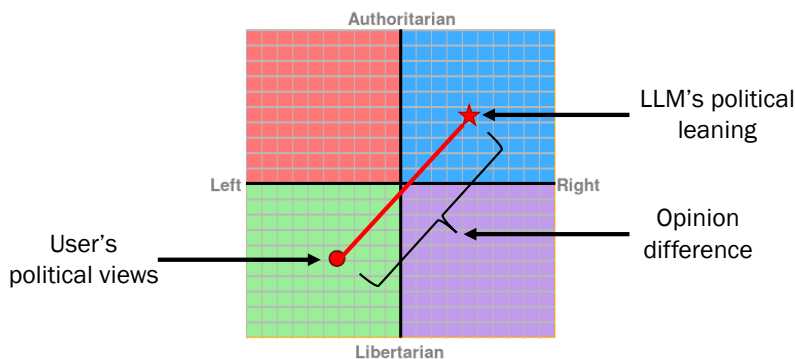

*Figure 4.*     *'Opinion difference' reflects the delta between the user's political viewpoint and the selected LLM's political leaning.*

It is important to note that 'neutrality' still cannot be purely objective given the subjective nature of the task at hand. In our context, neutrality is based on an operationalization using the political spectrum test and matrix. But what we often call 'neutral' is usually shaped by common social values or popular opinions, which can still be biased. In AI systems trained on large amounts of internet data, this kind of 'neutrality' might reflect the most common or dominant views and not a truly balanced perspective. Further, the "optimal" points in Figure 5 are estimations of the most effective solution for critical thinking depending on psychological effects of confirmation bias and cognitive dissonance. While not empirically validated, they serve as theoretical reference points and need to be tested in the future by psychologists.

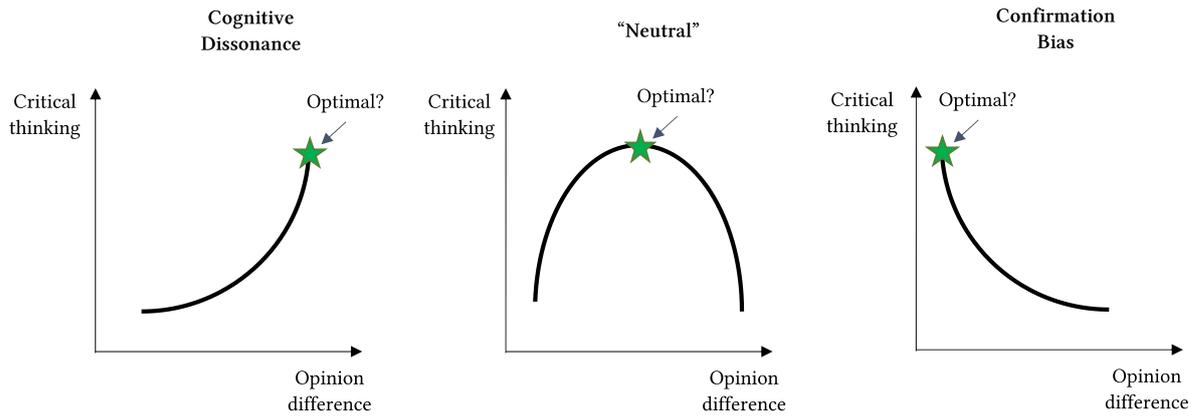

*Figure 5.    Scenarios to enhance critical thinking.*

When both the user and the LLM share similar political views (A-A), a low opinion difference leading to confirmation bias will be observed. This will result in the user agreeing with the tool's detection of propaganda, thereby enhancing their critical analysis of the article. Alternatively, when the user's and LLM's political views differ (A-B), a high opinion difference will create cognitive dissonance. This dissonance can either increase critical thinking as the user reconsiders their own beliefs in light of the tool's different perspective, or have no effect on it if the user distrusts and disengages with the tool. However, balancing these two strategies and finding the middle ground may be the most effective approach for enhancing critical thinking.

| User's political views | LLM's political leaning | Opinion difference | Cognitive effect | Potential effect on critical thinking |
|---|---|---|---|---|
| A | A | low | Confirmation bias | increased |
| A | B | high | Cognitive dissonance | increased or no effect |

*Table 1.    Possible potential effects on critical thinking.*

Transitioning from the theoretical exploration of scenarios that could enhance critical thinking, we now present two realization strategies formulated to translate these scenarios into user experiences. Our first strategy is to offer users the ability to choose an LLM with a political leaning that matches their preferences, ensuring they are aware of any model bias. This can be technically realized through clear disclaimers and an option to select from various politically labelled models (e.g., marking GPT-4 model as 'libertarian left' as suggested by (Feng et al., 2023)).

The second strategy involves personalizing the use of LLM to provide opinions that either: contrast with the user's views to induce cognitive dissonance, align with them to reinforce confirmation bias, or offer a 'neutral' option. Users have the option to self-reveal their political view or undergo a test for the system to determine it. Subsequently, the system either automatically selects an LLM with the corresponding political leaning or the LLM is prompted to adopt a specific viewpoint, such as *'Explain as if you were a model with authoritarian right-wing views.'* Table 2 summarizes the strategies and the proposed technical realization.

| Strategy | Description | Proposed technical realization |
|---|---|---|
| 1. Give users a choice to select a political leaning of an LLM to be used | (i) Make it transparent to users that the model is biased, and (ii) Let users select the political leaning of their preference. | (i) Disclaimer or FAQ, and (ii) Use an LLM with a political leaning of users' preference (e.g. provide selection from models labelling them as, e.g. 'left libertarian') or prompt the model to reflect the desired political view (e.g. 'Explain as if you were a model that has more authoritarian right-wing views.') |
| 2. Personalize based on political viewpoints to provide opposing opinion (to trigger cognitive dissonance), or confirmatory opinion (to trigger confirmation bias), or 'neutral' opinion. | (i) Identify users' political viewpoint (e.g. by self-disclosure or using any test that identifies political views), and (ii) Provide opinion opposing or confirmatory or 'neutral' to the users' political viewpoints. | (i) Ask the user to fill their political view in their profile or do a test to identify political views, and (ii) Use an LLM with a desired (contrasting / confirmatory / 'neutral') political leaning of users' preference or prompt the model to reflect the desired (contrasting / confirmatory / 'neutral') political view (e.g. 'Explain as if you were a model that has more authoritarian right-wing views.') |

*Table 2.        Strategies how to leverage inherent political biases and the proposed realization.*

# 5    Method

This study is exploratory in nature, aiming to understand users' viewpoints and concerns regarding inherent political biases in the explanations provided by Apollolytics. Specifically, we examine how users respond when these explanations align or misalign with their own political perspectives, as framed by the two proposed strategies. The main source of our empirical data was 20 semi-structured interviews, which were chosen as the data collection technique to facilitate rich, qualitative insights into participants' personal perspectives.

## 5.1    Participants

The interviews were conducted with participants of varying age, gender, and education, all residing in Germany (see Table 3), and recruited on Prolific. Prolific is a widely used platform for recruiting subjects for academic research because it provides access to a diverse and verified participant pool and is known for high-quality data and reliability (Peer et al., 2022). To recruit participants on Prolific, we established specific criteria: they must be native German-speakers as it was the main language of our study, not have a professional or academic background in political or communication sciences to represent 'average' news readers, and use a laptop or desktop, as Apollolytics' current implementation is not optimized for mobile devices. As an incentive for the participation and reward for their time and effort, the participants were paid 9 GBP / hour. The average age of the interview participants was between 25 and 44 years old, 7 females and 13 males. We initially invited 40 participants (18 male, 21 female and 1 with no information on gender) to test the Apollolytics extension, from which 20 agreed for a qualitative interview. While we ensured gender balance during the testing phase, the proportion shifted during the interviews, as participant selection at that stage was beyond our control. All participants signed a consent form in which they were also informed about the potential risks (such as influence of texts that might include propaganda) related to the study.

| No. | Pseudonym | Age Group | Gender | Occupation |
|---|---|---|---|---|
| 1 | Max | 18 to 24 | Male | Public administration student |
| 2 | Jack | 18 to 24 | Male | Intern in a non-profit organization |
| 3 | Sarah | 45 to 64 | Female | Daily companion |
| 4 | John | 25 to 44 | Male | SAP Consultant |
| 5 | Anna | 18 to 24 | Female | Sustainability student |
| 6 | Charles | 25 to 44 | Male | Pedagogy student |
| 7 | Peter | 45 to 64 | Male | Chemical laboratory technician |
| 8 | Paul | 25 to 44 | Male | Online marketing employee |
| 9 | Ted | 25 to 44 | Male | Business administration student |
| 10 | George | 25 to 44 | Male | Master student |
| 11 | Maria | 25 to 44 | Female | Researcher in biology |
| 12 | Sasha | 25 to 44 | Female | PhD student in psychology |
| 13 | Kate | 25 to 44 | Female | University administration |
| 14 | Mark | 18 to 25 | Male | Student of arts |
| 15 | Nick | 25 to 44 | Male | Engineer |
| 16 | Sonja | 25 to 44 | Female | Data Protection |
| 17 | Joseph | 25 to 44 | Male | Marketing |
| 18 | Klaus | 45 to 64 | Male | Engineer |
| 19 | Rudolf | 65+ | Male | Finance |
| 20 | Jessica | 25 to 44 | Female | Administration |

*Table 3.        Overview of the interviewees.*

## 5.2    Procedure

Before the interviews, the participants were asked to fill out a survey that included demographic questions. After that, the participants were presented with two different articles. The participants read the first article without any interaction with Apollolytics, representing the way we typically consume news today. For the second article, they interacted with Apollolytics while reading. Both articles were presented on webpages that mimicked online media outlets, created for the study (Figure 6).

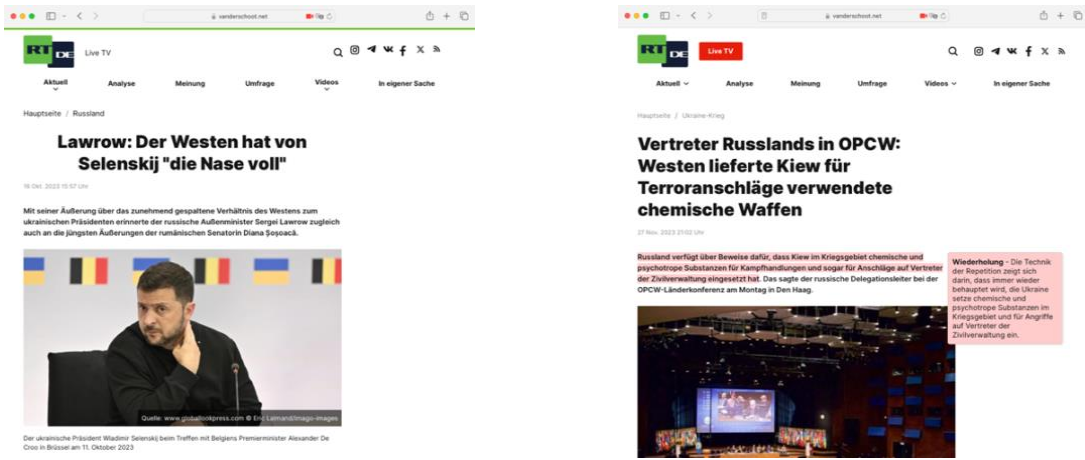

*Figure 6.        A mock webpage without (left) and with the interaction with Apollolytics (right).*

After that, we invited participants to share their most memorable positive and negative experiences with the propaganda detection tool and encouraged them to reflect on any challenges they encountered during its use. The interview in particular focused on the aspects of political biases in the explanations provided to users, users' perceptions and concerns about them, and their awareness of their own political leaning and bias. While the version of Apollolytics tested in the study used GPT-4 as its underlying model, participants did not have the ability to experiment with models of different political leanings. Nevertheless, we asked them to reflect on hypothetical scenarios in which the tool could provide explanations tailored to, or opposing, their political views. Their responses provide insight into perceptions and preferences. However, we also acknowledge a limitation here: future research is needed to observe actual user behaviour in an experimental settings to complement the self-reported attitudes. The interview questionnaire included the following questions:

- Did you notice any political biases in the extension's suggestions?
- How would you feel about the extension offering suggestions based on your personal political preferences?
- If the extension provided counter-opinions based on your political views, how would this affect your trust and experience?
- Would presenting alternative political viewpoints enhance or detract from your critical evaluation of content?
- If given the option to choose the political viewpoint the extension uses to analyze content, would you prefer it to align with your own views or present alternative opinions?

The total duration of the interviews was 594 minutes, ranging from 20 to 48 minutes per interview. The average interview length was 31 minutes. All interviews were conducted, recorded, and transcribed in an online conferencing tool (MS Teams). This study received ethical approval from an Institutional Review Board (IRB) under the ID OEC IRB #2023-076.

## 5.3  Data analysis and interpretation

We applied the reflexive thematic analysis (Braun & Clarke, 2021) to analyse and interpret our data. Our objective was to extract insights from participants' experiences and establish empirical categories grounded in their responses. The research team regularly met to interpret findings, reconcile differences, and find consensus on each thematic category. Throughout the process, we engaged in an iterative interpretation of the raw data and themes, identifying explicit and underlying patterns and relationships.

Initially, we immersed ourselves in the interview transcripts by thoroughly reading and re-reading them, noting initial observations and potential patterns. We then systematically coded the data, highlighting significant statements and assigning codes that captured the essence of those statements; this process resulted in a total of 18 initial codes, such as "perception of neutrality," "lack of immediate awareness of biases," "interest in alternative opinions," and "risk of echo chambers." Through collaborative discussions, we organized these 18 codes into three main themes based on conceptual similarities and relationships among participants' ideas. The first theme, **bias awareness**, encompassed codes related to participants' recognition (or lack thereof) of potential biases in the tool's explanations. The second theme, **strategies**, explored user preferences for personalization and choice within the tool and was further divided into two sub-themes: "personalization based on political viewpoints" and "choice." The third theme, **effect over time**, addressed how participants believe their trust and engagement with the tool may evolve over time. We reviewed and refined these themes to ensure they accurately represented the data and were distinct from each other, cross-checking them against the coded extracts and the entire data set. In the results section, we presented these themes, supported by direct quotes from participants to illustrate each point. Based on our analysis, we formulated three design recommendations (DR1, DR2, DR3) to guide the enhancement of the tool. By following this approach, we ensured that our analysis was thorough and grounded in the participants' actual responses.

# 6 Results and Design Recommendations

This section presents the interview results, organized into three main themes: participants' awareness of potential biases, two proposed strategies - personalization and choice - and their potential long-term effects. In each, we formulate a design recommendation (DR).

## 6.1 Bias awareness

This theme explores users' awareness of potential inherent biases in the explanations generated by the tool. Participants generally perceive themselves as critical thinkers. However, most confess a lack of immediate awareness regarding potential biases in explanations provided by the tool. The current format of explanations is perceived as neutral enough, as noted by Sarah. However, participants like Kate and Jack suggested that their failure to detect model bias in confirmatory opinions might be due to these opinions aligning with their own libertarian views. The perception of bias could differ if confronted with contrary opinions. However, when biases are explicitly pointed out and discussed in the interview, our participants showed increased engagement in the discussion. At this point, the participants agreed that the tool has quite noticeable 'western' bias. Altogether, this suggests an underestimation or unawareness of inherent biases in the tool, with this, we derive the following design recommendation:

**DR1. Strengthen bias awareness**: Incorporate features that explicitly highlight biases in the tool's explanations, facilitating user awareness and engagement in critical discussion about these biases. Examples of such tools could be bias dashboards, explanatory features like the one presented in Apollolytics, and user personalization. Such interventions can also help counter the formation of echo chambers, where users are primarily exposed to viewpoints that reinforce their existing beliefs. By making biases explicit and by incorporating bias transparency in the tool's design, users are more likely to critically evaluate their own beliefs (confirmation bias) and become more aware of differing viewpoints, potentially reducing cognitive dissonance.

## 6.2 Strategies

Another recurring theme was possible *strategies* to address the inherited model biases. While *personalisation* enables users to tailor the political perspective of the tool's explanations, providing *choice* allows for greater flexibility and the ability to switch between different political viewpoints.

**Personalization based on political viewpoints**. This theme explores user preferences for personalization in the tool and the types of opinions it should display. Initially, there was skepticism among users about personalization, with a strong preference for a neutral approach to avoid feeling manipulated. Yet, when discussing the option of presenting alternative opinions, some participants, like John, saw it as beneficial and intriguing, as illustrated by his comment: "*That would be actually very interesting tool. Funny. I would enjoy it*". Nonetheless, no one expressed interest in a version of the tool that would provide only opposing viewpoints. Most of the users would immediately stop using the tool or consider it "propagandistic" itself. For example, Charles formulates it like this: "*If it's very different, I would become suspicious.*"

Maria and Kate raised concern that such tool could be "propagandistic" itself and it would be nice to have more transparency on it. On top of that they mentioned that occasionally it would be interesting to 'switch' to an opposite political mode to be more prepared for the argumentation with the people of opposite political views in their personal or professional network. Furthermore, some of the participants (Joseph, Klaus) saw a risk of increasing echo chambers effect. Thus, the possibility to have opposing viewpoints serves as an effective safeguard against the 'strong radicalization' of people's opinion (Klaus).

Transparency in personalization emerged as a key concern, e.g. for George and Sasha, highlighting the need for users to understand the development and operation of the tool, as well as its potential political biases, which could then be adjusted in the settings. This need for transparency also ties into the concern about the potential misuse of the tool, as there is a risk that biases could be exploited for manipulation.

**Choice**. Talking about preferences for having a choice of the participants, most of them would prefer explanations either to align with their already existing views or to choose the neutral point of view.

Some participants (like Max, Anna and Maria) who are more into reading news, found it interesting to investigate the opposite viewpoints to get a wider perspective on the news in order to critically assess their own stance from time to time. Anna mentioned that her preferred mode will be neutral as her interest from such a tool is to get as objective information as possible. For Max, such features would help him understand where his media stands on the political spectrum. Furthermore, there is also interest in opposing viewpoints to critically assess one's 'bubble', as noted by Max claiming that true objectivity requires exposure to diverse viewpoints: "*To see your bubble, your interests a bit more critical [with the help of the tool]*". Considering this, we formulate the following design recommendation:

**DR2. Enable flexible personalization and choice**: Offer users the ability to personalize the political viewpoint of the tool's explanations. Maintain a default neutral setting - meaning the tool uses an LLM that provides detections explanations without favouring any political side, based on centrist viewpoints (based on the political spectrum) - while allowing users to explore alternative or confirmatory viewpoints as desired. This allows users to confront or align with their biases. Choosing confirmatory views reinforces confirmation bias, while exploring opposing views can induce cognitive dissonance, leading to more critical thinking.

## 6.3    Effect over time

This theme described users' perspectives on how their trust and engagement with the tool may evolve over time. Initially presenting opposing views can lead to distrust; it is important to first build user confidence before gradually introducing alternative perspectives. Participants, especially those like Max or Anna who closely follow the news and are interested in politics, expressed interest in occasionally exploring opposing views for a broader understanding or better argumentation, recognizing that all sources have some bias: "*I would for sure try it out and look into different perspectives.*" Over time, as users become more adept at recognizing propaganda techniques, their reliance on the tool may diminish, as they may no longer need it to identify different viewpoints on their own, as suggested by Ted. Thus, we derive the following design recommendation:

**DR3. Gradually introduce diverse perspectives**: To build trust, initially align the tool's output with user views, gradually introducing diverse perspectives over time to expand user understanding and reduce reliance on the tool as users become more adept at identifying propaganda techniques. This design recommendation addresses cognitive dissonance by slowly exposing users to alternative viewpoints, reducing the initial discomfort or rejection, and encouraging a better and more critical understanding over time.

## 7    Discussion and Conclusion, Limitations

This exploratory work set out to answer the following research question, '*How can we leverage the inherent political biases of LLMs in the design of AI propaganda detection tools to enhance users' critical thinking?*' We propose two strategies based on psychological cognitive effects, such as confirmation bias and cognitive dissonance, and gathered users' perceptions through interviews. The study found that participants were often unaware of inherent biases in AI-generated explanations unless explicitly pointed out, suggesting that bias-awareness features are crucial for encouraging users to critically reflect on content. By offering personalization options that align or contrast with users' political views, AI tools can trigger confirmation bias or cognitive dissonance, both of which can stimulate critical thinking. Gradual exposure to diverse viewpoints over time was a way to build trust and deepen users' critical engagement, particularly as they become more adept at recognizing biases on their own.

However, the findings must be considered alongside certain ethical implications. Aligning AI explanations with users' political views might reinforce existing echo chambers (Dahlgren, 2020; Dubois et al., 2020) and contribute to polarization. On the other hand, exposing users to conflicting views too abruptly may erode trust or lead to disengagement, particularly for users with pre-existing skepticism toward AI systems. For example, participants in our study raised concerns that overly one-sided outputs, especially when not clearly explained, could make the tool "propagandistic" itself.

To mitigate these risks, future designs of tools like Apollolytics may incorporate further mechanisms for transparency and user control. This includes customizable bias settings, clear disclosures about model leanings, visualizations of bias through dashboards, and contextual explanations to help users understand how and why outputs are generated. For instance, users could be informed if a model reflects a particular ideological stance and have the option to adjust or switch between models.

The findings suggest that leveraging LLM biases in a transparent and flexible manner, while allowing users to explore multiple perspectives, can enhance their ability to critically assess news content. The proposed strategies (bias awareness, personalization and choice, and gradual introduction of diverse perspectives) could be applied to a variety of AI tools beyond propaganda detection, such as news recommendation systems or social media platforms, to foster critical thinking in diverse user groups. The generalizability of these strategies may extend to different cultural and political contexts, as long as transparency and user control over bias exposure are maintained. Further research is needed to identify the types of applications and explore how other kinds of biases, such as gender and racial biases, can be leveraged in AI tools, and when it will no longer be appropriate.

While in our study, we offer users a design that would create a shift from irrational unconscious behavior to more deliberate, we acknowledge that in practice, emotional responses, varying levels of digital literacy, and pre-existing distrust in AI systems could still hinder engagement with the tool. For example, users encountering cognitive dissonance may disengage entirely rather than re-evaluate their beliefs. Incorporating features like empathetic explanations, interactive tutorials, or adaptive designs that account for diverse user behaviours could enhance the tool's effectiveness and ensure broader usability. This calls for more experimental research to test the outcomes of the interactions with the tool.

We also acknowledge the following limitations, which may also inspire future work. First, the study relies solely on qualitative interviews, which may not fully capture complex cognitive processes like cognitive dissonance and confirmation bias. In the future, this can be addressed by combining interviews with quantitative methods such as experiments or eye-tracking to empirically assess whether these cognitive effects occur during interactions with AI tools. Since the study's exploratory nature limits its ability to definitively measure cognitive biases, future research could use experimental designs that manipulate political alignment and assess changes in beliefs or trust, offering a clearer understanding of how confirmation bias and cognitive dissonance impact user interactions with AI. Second, the study focuses on a single propaganda detection tool instead of a class of such tools. This could be mitigated in future research by exploring how different AI tools across various contexts, such as social media or news platforms, influence users' critical thinking. Third, since our study does not give any insight into long-term effects of the use and potential unintended consequences, such as over-reliance or misuse, these should be studied further. Another limitation is the geographic and cultural scope of our research. The study's focus on a Western, German-speaking context means that the results and strategies may not generalize to other regions, where political spectrums and propaganda techniques may differ. Political leaning is not a universal axis, and users from non-Western or multilingual contexts may interpret neutrality, bias, or propaganda differently. Finally, a crucial aspect of such research is to further explore the ethical implications of cognitive manipulation through such AI tools in order not to harm potential users.

# 8    Acknowledgements

We thank the Digitalization Initiative of the Zurich Higher Education Institutions (DIZH) for financing this study under the DIZH Founder Call. We also acknowledge the use of generative AI (Grammarly and ChatGPT) for text editing.

# References

Abid, A., Farooqi, M., & Zou, J. (2021). Persistent anti-muslim bias in large language models. *Proceedings of the 2021 AAAI/ACM Conference on AI, Ethics, and Society*, 298–306.

Ahmed, A. A. A., Aljabouh, A., Donepudi, P. K., & Choi, M. S. (2021). Detecting fake news using machine learning: A systematic literature review. *arXiv Preprint arXiv:2102.04458*.

Akerlof, G. A., & Dickens, W. T. (1982). The Economic Consequences of Cognitive Dissonance. *The American Economic Review*, 72(3), 307–319.

Barrón-Cedeno, A., Jaradat, I., Da San Martino, G., & Nakov, P. (2019). Proppy: Organizing the news based on their propagandistic content. *Information Processing & Management*, 56(5), 1849–1864.

Bender, E. M., Gebru, T., McMillan-Major, A., & Shmitchell, S. (2021). On the Dangers of Stochastic Parrots: Can Language Models Be Too Big? 🦜. *Proceedings of the 2021 ACM Conference on Fairness, Accountability, and Transparency*, 610–623.

Bommasani, R., Hudson, D. A., Adeli, E., Altman, R., Arora, S., von Arx, S., Bernstein, M. S., Bohg, J., Bosselut, A., Brunskill, E., & others. (2021). On the opportunities and risks of foundation models. *arXiv Preprint arXiv:2108.07258*.

Boonprakong, N., Chen, X., Davey, C., Tag, B., & Dingler, T. (2023). Bias-Aware Systems: Exploring Indicators for the Occurrences of Cognitive Biases when Facing Different Opinions. *Proceedings of the 2023 CHI Conference on Human Factors in Computing Systems*, 1–19. https://doi.org/10.1145/3544548.3580917

Boonprakong, N., Tag, B., & Dingler, T. (2023). Designing Technologies to Support Critical Thinking in an Age of Misinformation. *IEEE Pervasive Computing*. https://ieeexplore.ieee.org/abstract/document/10142163/

Bran, A., & Vaidis, D. C. (2020). On the Characteristics of the Cognitive Dissonance State: Exploration Within the Pleasure Arousal Dominance Model. *Psychologica Belgica*, 60(1), 86. https://doi.org/10.5334/pb.517

Braun, V., & Clarke, V. (2021). *Thematic Analysis: A Practical Guide*. SAGE Publications Ltd.

Brown, T., Mann, B., Ryder, N., Subbiah, M., Kaplan, J. D., Dhariwal, P., Neelakantan, A., Shyam, P., Sastry, G., Askell, A., & others. (2020). Language models are few-shot learners. *Advances in Neural Information Processing Systems*, 33, 1877–1901.

Da San Martino, G., Cresci, S., Barrón-Cedeño, A., Yu, S., Di Pietro, R., & Nakov, P. (2020). A survey on computational propaganda detection. *arXiv Preprint arXiv:2007.08024*.

Da San Martino, G., Shaar, S., Zhang, Y., Yu, S., Barrón-Cedeno, A., & Nakov, P. (2020). Prta: A system to support the analysis of propaganda techniques in the news. *Proceedings of the 58th Annual Meeting of the Association for Computational Linguistics: System Demonstrations*, 287–293.

Dahlgren, P. M. (2020). *Media Echo Chambers: Selective Exposure and Confirmation Bias in Media Use, and its Consequences for Political Polarization*. https://gupea.ub.gu.se/handle/2077/67023

Dastin, J. (2022). Amazon scraps secret AI recruiting tool that showed bias against women. In *Ethics of data and analytics* (pp. 296–299). Auerbach Publications. https://books.google.ch/books?hl=en&lr=&id=E51kEAAAQBAJ&oi=fnd&pg=PA296&dq=Dastin,+J.+(2022).+Amazon+Scraps+Secret+AI+Recruiting+Tool+That+Showed+Bias+against+Women*.+In+Ethics+of+Data+and+Analytic&ots=INiAdZABI-&sig=7wryGBeF6iSsoY3I00GxRvVrAwE

Della Vedova, M. L., Tacchini, E., Moret, S., Ballarin, G., DiPierro, M., & De Alfaro, L. (2018). Automatic online fake news detection combining content and social signals. *2018 22nd Conference of Open Innovations Association (FRUCT)*, 272–279.

Devlin, J., Chang, M.-W., Lee, K., & Toutanova, K. (2019). BERT: Pre-training of Deep Bidirectional Transformers for Language Understanding. *Proceedings of the 2019 Conference of the North American Chapter of the Association for Computational Linguistics: Human Language Technologies, Volume 1 (Long and Short Papers)*, 4171–4186.

Dubois, E., Minaeian, S., Paquet-Labelle, A., & Beaudry, S. (2020). Who to Trust on Social Media: How Opinion Leaders and Seekers Avoid Disinformation and Echo Chambers. *Social Media + Society*, 6(2), 2056305120913993. https://doi.org/10.1177/2056305120913993

Feng, S., Park, C. Y., Liu, Y., & Tsvetkov, Y. (2023). From Pretraining Data to Language Models to Downstream Tasks: Tracking the Trails of Political Biases Leading to Unfair NLP Models. *arXiv Preprint arXiv:2305.08283*.

Gallegos, I. O., Rossi, R. A., Barrow, J., Tanjim, M. M., Kim, S., Dernoncourt, F., Yu, T., Zhang, R., & Ahmed, N. K. (2023). Bias and fairness in large language models: A survey. *arXiv Preprint arXiv:2309.00770*.

Harmon-Jones, E., & Mills, J. (2019). An introduction to cognitive dissonance theory and an overview of current perspectives on the theory. In *Cognitive dissonance: Reexamining a pivotal theory in psychology, 2nd ed* (pp. 3–24). American Psychological Association. https://doi.org/10.1037/0000135-001

Hasanain, M., Ahmed, F., & Alam, F. (2024). *Large Language Models for Propaganda Span Annotation* (arXiv:2311.09812). arXiv. http://arxiv.org/abs/2311.09812

Hoferer, N., Sprenkamp, K., Quelle, D. C., Jones, D. G., Katashinskaya, Z., Bovet, A., & Zavolokina, L. (2025). Effective Yet Ephemeral Propaganda Defense: There Needs to Be More than One-Shot Inoculation to Enhance Critical Thinking. *Extended Abstracts of the CHI Conference on Human Factors in Computing Systems (CHI EA '25), April 26-May 1, 2025, Yokohama, Japan. ACM, New York, NY, USA, 13 Pages.* https://doi.org/10.1145/3706599.3720125

Im, C., Shin, Y., Lee, S., & Kim, J. (2014). A research of design principles to reduce cognitive load in online learning platform. *Proceedings of HCI Korea*, 306–313.

Ji, K. (2023). Quantifying and Measuring Confirmation Bias in Information Retrieval Using Sensors. *Adjunct Proceedings of the 2023 ACM International Joint Conference on Pervasive and Ubiquitous Computing & the 2023 ACM International Symposium on Wearable Computing*, 236–240. https://doi.org/10.1145/3594739.3610765

Kahneman, D. (2011). *Thinking, fast and slow*. macmillan.

Kleinberg, J., Mullainathan, S., & Raghavan, M. (2016). *Inherent Trade-Offs in the Fair Determination of Risk Scores* (arXiv:1609.05807). arXiv. http://arxiv.org/abs/1609.05807

Levin, D., Harriott, C., Paul, N. A., Zheng, T., & Adams, J. A. (2013). Cognitive Dissonance as a Measure of Reactions to Human-Robot Interaction. *Journal of Human-Robot Interaction*, *2*(3), 1–17. https://doi.org/10.5898/JHRI.2.3.Levin

Levy, G., & Razin, R. (2019). Echo chambers and their effects on economic and political outcomes. *Annual Review of Economics*, *11*(1), 303–328.

Liu, Y., Ott, M., Goyal, N., Du, J., Joshi, M., Chen, D., Levy, O., Lewis, M., Zettlemoyer, L., & Stoyanov, V. (2019). Roberta: A robustly optimized bert pretraining approach. *arXiv Preprint arXiv:1907.11692*.

Lu, K., Mardziel, P., Wu, F., Amancharla, P., & Datta, A. (2020). Gender bias in neural natural language processing. *Logic, Language, and Security: Essays Dedicated to Andre Scedrov on the Occasion of His 65th Birthday*, 189–202.

Mehrabi, N., Morstatter, F., Saxena, N., Lerman, K., & Galstyan, A. (2021). A survey on bias and fairness in machine learning. *ACM Computing Surveys (CSUR)*, *54*(6), 1–35.

Min, A. (2023). Artifical Intelligence and Bias: Challenges, Implications, and Remedies. *Journal of Social Research*, *2*, 3808–3817. https://doi.org/10.55324/josr.v2i11.1477

Nayak, S., Baumann, T., Bhattacharya, S., Karakanta, A., Negri, M., & Turchi, M. (2021). See me Speaking? Differentiating on Whether Words are Spoken On Screen or Off to Optimize Machine Dubbing. *Companion Publication of the 2020 International Conference on Multimodal Interaction*, 130–134. https://doi.org/10.1145/3395035.3425640

Ohagi, M. (2024). Polarization of autonomous generative AI agents under echo chambers. *arXiv Preprint arXiv:2402.12212*.

Olsen, R. A. (2008). Cognitive Dissonance: The Problem Facing Behavioral Finance. *Journal of Behavioral Finance*, *9*(1), 1–4. https://doi.org/10.1080/15427560801896552

OpenAI. (2023). *GPT-4 Technical Report*. https://cdn.openai.com/papers/gpt-4.pdf

Patil, M., Yadav, H., Gawali, M., Suryawanshi, J., Patil, J., Yeole, A., Shetty, P., & Potlabattini, J. (2024). A Novel Approach to Fake News Detection Using Generative AI. *International Journal of Intelligent Systems and Applications in Engineering*, *12*(4s), 343–354.

Peer, E., Rothschild, D., Gordon, A., Evernden, Z., & Damer, E. (2022). Data quality of platforms and panels for online behavioral research. *Behavior Research Methods*, 1.

Piskorski, J., Stefanovitch, N., Da San Martino, G., & Nakov, P. (2023). Semeval-2023 task 3: Detecting the category, the framing, and the persuasion techniques in online news in a multi-lingual setup. *Proceedings of the 17th International Workshop on Semantic Evaluation (SemEval-2023)*, 2343–2361.

Qawasmeh, E., Tawalbeh, M., & Abdullah, M. (2019). Automatic identification of fake news using deep learning. *2019 Sixth International Conference on Social Networks Analysis, Management and Security (SNAMS)*, 383–388.

Qian, Y., Muaz, U., Zhang, B., & Hyun, J. W. (2019). Reducing gender bias in word-level language models with a gender-equalizing loss function. *arXiv Preprint arXiv:1905.12801*.

Quattrociocchi, W., Scala, A., & Sunstein, C. R. (2016). Echo chambers on Facebook. *Available at SSRN 2795110*.

Radford, A., Wu, J., Child, R., Luan, D., Amodei, D., Sutskever, I., & others. (2019). Language models are unsupervised multitask learners. *OpenAI Blog*, *1*(8), 9.

Richter, T., & Maier, J. (2017). Comprehension of Multiple Documents With Conflicting Information: A Two-Step Model of Validation. *Educational Psychologist*, *52*(3), 148–166. https://doi.org/10.1080/00461520.2017.1322968

Rozado, D. (2023). The political biases of ChatGPT. *Social Sciences*, *12*(3), 148.

Ryan-Mosleyarchive, T. (2023). *How generative AI is boosting the spread of disinformation and propaganda*. MIT Technology Review. https://www.technologyreview.com/2023/10/04/1080801/generative-ai-boosting-disinformation-and-propaganda-freedom-house/

Schwind, C., Buder, J., & Hesse, F. W. (2011). I will do it, but i don't like it: User reactions to preference-inconsistent recommendations. *Proceedings of the SIGCHI Conference on Human Factors in Computing Systems*, 349–352. https://doi.org/10.1145/1978942.1978992

Sheng, E., Chang, K.-W., Natarajan, P., & Peng, N. (2020). " nice try, kiddo": Investigating ad hominems in dialogue responses. *arXiv Preprint arXiv:2010.12820*.

Sprenkamp, K., Jones, D. G., & Zavolokina, L. (2023). *Large Language Models for Propaganda Detection* (arXiv:2310.06422). arXiv. https://doi.org/10.48550/arXiv.2310.06422

Telci, E. E., Maden, C., & Kantur, D. (2011). The theory of cognitive dissonance: A marketing and management perspective. *Procedia - Social and Behavioral Sciences*, *24*, 378–386. https://doi.org/10.1016/j.sbspro.2011.09.120

Thornhill, C., Meeus, Q., Peperkamp, J., & Berendt, B. (2019). A digital nudge to counter confirmation bias. *Frontiers in Big Data*, *2*, 11.

Touvron, H., Lavril, T., Izacard, G., Martinet, X., Lachaux, M.-A., Lacroix, T., Rozière, B., Goyal, N., Hambro, E., Azhar, F., & others. (2023). Llama: Open and efficient foundation language models. *arXiv Preprint arXiv:2302.13971*.

Tversky, A., & Kahneman, D. (1974). Judgment under uncertainty: Heuristics and biases. *Science*, *185*(4157), 1124–1131.

Walton, D. (1997). What is propaganda, and what exactly is wrong with it. *Public Affairs Quarterly*, *11*(4), 383–413.

Wang, X., Ge, T., Mao, A., Li, Y., Wei, F., & Chen, S.-Q. (2022). Pay attention to your tone: Introducing a new dataset for polite language rewrite. *arXiv Preprint arXiv:2212.10190*.

Xin, D., Mayoraz, N., Pham, H., Lakshmanan, K., & Anderson, J. R. (2017). Folding: Why Good Models Sometimes Make Spurious Recommendations. *Proceedings of the Eleventh ACM Conference on Recommender Systems*, 201–209. https://doi.org/10.1145/3109859.3109911

Zavolokina, L., Sprenkamp, K., Katashinskaya, Z., Jones, D. G., & Schwabe, G. (2024). Think Fast, Think Slow, Think Critical: Designing an Automated Propaganda Detection Tool. *Proceedings of the CHI Conference on Human Factors in Computing Systems*, 1–24. https://doi.org/10.1145/3613904.3642805

Zhang, B. H., Lemoine, B., & Mitchell, M. (2018). Mitigating unwanted biases with adversarial learning. *Proceedings of the 2018 AAAI/ACM Conference on AI, Ethics, and Society*, 335–340.